# BOTTOMONIUM AND BOTTOMONIUM-LIKE STATES AND DECAYS AT BELLE

G. TATISHVILI [a]

*Pacific Northwest National Laboratory*
*902 Battelle Boulevard, Richland, WA, USA*

Recent results from the Belle experiment are presented. We report the results of the first observation of P-wave spin-singlet Bottomonium states, observation of two charged Bottomonium-like resonances and the first observation of the radiative transition $h_b(1P) \rightarrow \eta_b(1S)\gamma$ at the $\Upsilon(5S)$ resonance region.

We report the results of the first observation of P-wave spin-singlet Bottomonium states $h_b(1P)$ and $h_b(2P)$ produced in the $\Upsilon(5S)$ region. We used a 121.4 fb$^{-1}$ data sample collected near the peak of the $\Upsilon(5S)$ resonance with the Belle detector at the KEKB asymmetric-energy $e^+e^-$ collider. The Belle detector is a large-solid-angle magnetic spectrometer consisting of a central drift chamber, an array of aerogel threshold Cherenkov counters, electromagnetic calorimeter composed of CsI(Tl) crystals located inside a superconducting solenoid with 1.5 T magnetic field. The detector is described in detail elsewhere [1].

The $h_b(nP)$ states were produced via $e^+e^- \rightarrow h_b(nP)\pi^+\pi^-$ and observed in the $\pi^+\pi^-$ missing mass spectrum of hadronic events [2]. The $\pi^+\pi^-$ missing mass was calculated by formula $M^2_{miss}=(P_{\Upsilon(5S)} - P_{\pi^+\pi^-})^2$, where $P_{\Upsilon(5S)}$ (4-momentum of the $\Upsilon(5S)$) was determined from the beam momenta. $P_{\pi^+\pi^-}$ is the 4-momentum of the $\pi^+\pi^-$ system. The background subtracted inclusive $M_{miss}$ spectrum is presented in Fig.1. To determine the number of produced resonant decays the $M_{miss}$ spectrum was fitted separately into three adjacent regions. The measured masses of $h_b(1P)$ and $h_b(2P)$ states are $M=(9898.2^{+1.1+1.0}_{-1.0-1.1})$ MeV/c$^2$ and $M=(10259.8\pm0.6^{+1.4}_{-1.0})$ MeV/c$^2$, respectively. Using the measured $h_b$ and world average masses of $\chi_{bJ}(nP)$ states we determine the hyperfine splittings to be $\Delta M_{\text{HF}} = (+1.7 \pm 1.5)$ and $(+0.5 ^{+1.6}_{-1.2})$ MeV/c$^2$ for $h_b(1P)$ and $h_b(2P)$, respectively which are consistent with zero. We measured the ratio of cross sections $R = \sigma(h_b(nP)\pi^+\pi^-) / \sigma(\Upsilon(2S)\pi^+\pi^-)$ to be $0.45\pm0.08 ^{+0.07}_{-0.12}$ for the $h_b(1P)$ and $0.77\pm0.08 ^{+0.22}_{-0.17}$ for the $h_b(2P)$, which indicates that $h_b(nP)\pi^+\pi^-$ and $\Upsilon(2S)\pi^+\pi^-$ proceed at similar rates despite the fact that the production of $h_b(nP)$ requires a spin flip of a $b$ quark. The angular analysis of the $\Upsilon(5S) \to h_b(1P)\pi^+\pi^-$ transition indicates spin parity of $J^P=1^+$ for the $h_b(1P)$ state. We also analyzed 711fb$^{-1}$ data at the $\Upsilon(4S)$ resonance to search for $h_b(1P)\pi^+\pi^-$ transition and set an upper limit on the ratio of $\sigma(e^+e^- \to h_b(1P)\pi^+\pi^-)$ at the $\Upsilon(4S)$ to that at the $\Upsilon(5S)$ of 0.27 at 90% C.L.

---

[a] On behalf of the Belle Collaboration

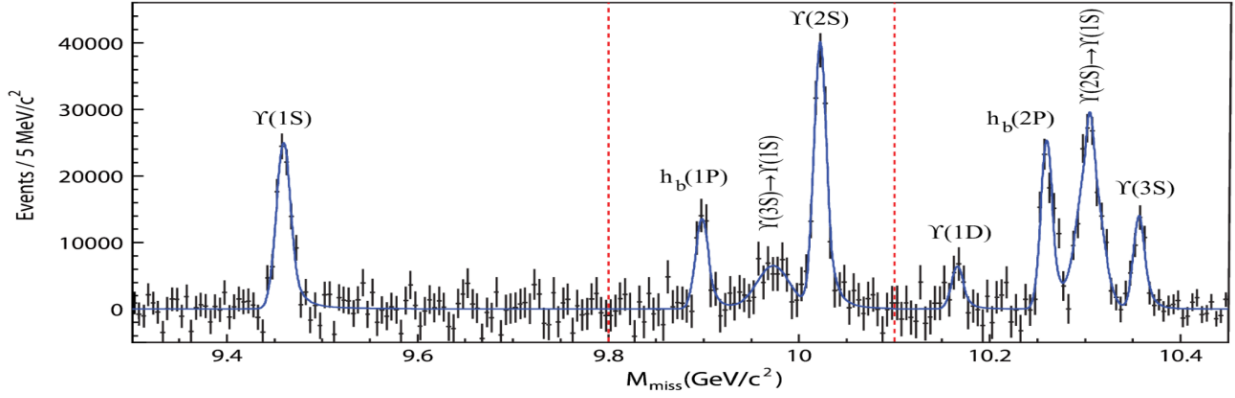

FIG. 1 The background subtracted inclusive $M_{\text{miss}}$ spectrum (points with errors). The vertical lines indicate boundaries of the fit regions. Overlaid smooth curve is the resulting fit function.

The analysis of di-pion transition of $\Upsilon(5S)$ resonance shows a high rates of $\Upsilon(5S) \to \Upsilon(nS)\pi^+\pi^-$ (n=1,2,3) and $\Upsilon(5S) \to h_b(mP)\pi^+\pi^-$ (m=1,2) which indicates contribution of exotic mechanisms in the $\Upsilon(5S)$ decays. We report results of study of resonant substructure in the decays of $\Upsilon(5S) \to \Upsilon(nS)\pi^+\pi^-$ (n=1,2,3) and $\Upsilon(5S) \to h_b(mP)\pi^+\pi^-$ (m=1,2) [3]. Fig.2 (a) and 2(b) show Dalitz distributions – the maximum value of the two $M^2[\Upsilon(2S)\pi^\pm]$ versus $M^2(\pi^+\pi^-)$ for $\Upsilon(2S)$ sideband and signal regions, respectively. Two horizontal bands in Fig. 2(b) indicate existence of structures in $\Upsilon(nS)\pi$ system near 10.61 GeV/c$^2$ ($Z_b(10610)$) and 10.65 GeV/c$^2$ ($Z_b(10650)$). Fig.3 show the yield of $\Upsilon(5S) \to h_b(mP)\pi^+\pi^-$ (m=1,2) as a function of pion missing mass. A clear two-peak structure indicates the production of $Z_b(10610)$ and $Z_b(10650)$. In total we observed two $Z_b(10610)$ and $Z_b(10650)$ Bottomonium-like resonances in five different decay channels $\Upsilon(nS)\pi^\pm$ (n=1,2,3) and $h_b(mP)\pi^\pm$ (m=1,2). The minimal quark content of the $Z_b(10610)$ and $Z_b(10650)$ requires a four quark combination. Weighted average values of masses and widths over all five channels are $M=10607.2\pm2.0$ MeV/c$^2$, $\Gamma=18.4\pm2.4$ MeV for the $Z_b(10610)$ and $M=10652.2\pm1.5$ MeV/c$^2$, $\Gamma=11.5\pm2.2$ MeV for the $Z_b(10650)$. The measured masses of these states are a few MeV/c$^2$ above the thresholds for the open beauty channels which suggests that their internal dynamics is dominated by the coupling to B meson pairs [4].

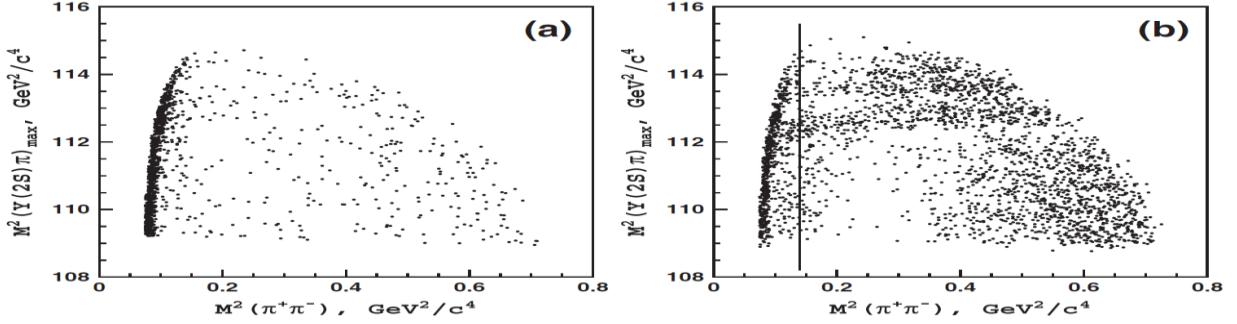

FIG. 2 Dalitz plots for the Υ(2S) sideband (a) and Υ(2S) signal (b) regions. Vertical line shows the cut which allows to remove a background from photon conversions in the detector elements.

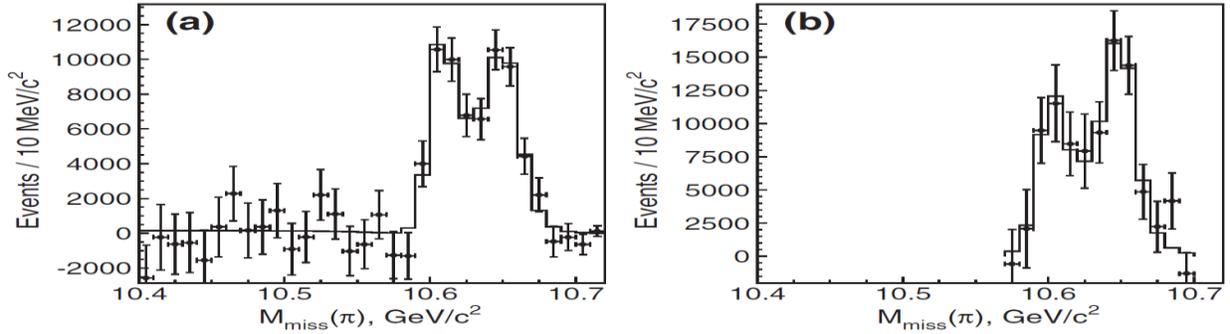

FIG. 3. The $h_b(1P)$ (a) and $h_b(2P)$ (b) yields as a function of the missing mass recoiling against the pion. Fit results are presented as a histograms.

The radiative transition to the $\eta_b(1S)$ is expected to be one of the dominant decay modes of the $h_b(1P)$. We report the first observation of the radiative transition $h_b(1P) \rightarrow \eta_b(1S)\gamma$ [5]. In the decay chain $\Upsilon(5S) \rightarrow Z^+_b \pi^-$, $Z^+_b \rightarrow h_b(1P)\pi+$, $h_b(1P) \rightarrow \eta_b(1S)\gamma$ we reconstruct only the π−, π+ and γ. We search for the $\eta_b(1S)$ signal in the distribution of $\Delta M_{\text{miss}}(\pi^+\pi^-\gamma) - M_{\text{miss}}(\pi^+\pi^-) + m[h_b(1P)]$ (see Fig.4). Observed signal parameterized by a non-relativistic Breit-Wigner function, the combinatorial background – by an exponential function.

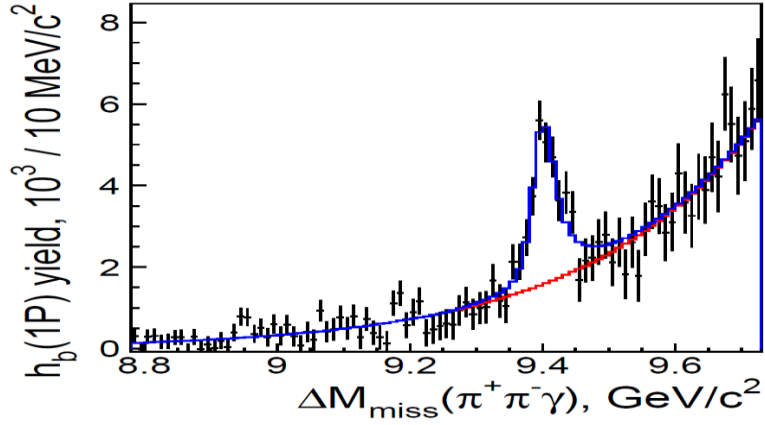

FIG. 4 $\Delta M_{miss}(\pi^+\pi^-\gamma)$ distribution of the $h_b(1P)$ yield with fit result superimposed.

We obtained the single most precise measurement of the $\eta_b(1S)$ mass, $(9401.0\pm1.9^{+1.4}_{-2.4})$ MeV/c$^2$, which corresponds to the hyperfine splitting $\Delta M_{HF}[\eta_b(1S)] = (59.3\pm1.9^{+2.4}_{-1.4})$ MeV/c$^2$. We report the first measurement of the $\eta_b(1S)$ width ($12.4^{+5.5+11.5}_{-4.6-3.4}$) MeV. For the branching fraction we find $B[h_b(1P) \to \eta_b(1S)\gamma] = (49.8\pm6.8^{+10.9}_{-5.2})$ % which agrees with the theoretical expectations [6].

## Acknowledgments

We thank the KEKB group for excellent operation of the accelerator; the KEK cryogenics group for efficient solenoid operations; the KEK computer group and the NII for valuable computing and SINET4 network support. We acknowledge support from MEXT, JSPS and Nagoya's TLPRC (Japan); ARC and DIISR (Australia); NSFC (China); MSMT (Czechia); DST (India); MEST, NRF, NSDC of KISTI, and WCU (Korea); MNiSW (Poland); MES and RFAAE (Russia); ARRS (Slovenia); SNSF (Switzerland); NSC and MOE (Taiwan); and DOE and NSF (USA).